\documentclass[twocolumn,aps,prl,showpacs,preprintnumbers,bibnotes,amsmath,amssymb,dvips]{revtex4}

\usepackage{natbib}
\usepackage{graphicx}
\usepackage{bm}
\usepackage{color}

\newcommand\unit[2]{\ensuremath{#1~\mathrm{{#2}}}}

\newcommand\SSZ{\ensuremath{{^1}\mathrm{S}_0}}
\newcommand\SPO{\ensuremath{{^1}\mathrm{P}_1}}

\newcommand\TPO{\ensuremath{{^3}\mathrm{P}_1}}

\renewcommand{\vec}[1]{\ensuremath{\bm{#1}}}

\begin{document}

\title{Systematic study of Optical Feshbach Resonances in an ideal gas}

\author{S. Blatt$^1$}
\author{T. L. Nicholson$^1$}
\author{B. J. Bloom$^1$}
\author{J. R. Williams$^1$}
\author{J. W. Thomsen$^{1}$}
\altaffiliation{
 Permanent address: The Niels Bohr Institute, Universitetsparken 5, 2100
 Copenhagen, Denmark.}
\author{P. S. Julienne$^2$}
\author{J. Ye$^1$}
\affiliation{$^1$JILA and Department of Physics, NIST and
  University of Colorado, Boulder, CO
  80309-0440, USA \\
  $^2$Joint Quantum Institute, NIST and the University of
  Maryland, Gaithersburg, MD 20899-8423,
  USA.}

\date{June 3, 2011}

\begin{abstract}
  Using a narrow intercombination line in alkaline earth atoms to
  mitigate large inelastic losses, we explore the Optical Feshbach
  Resonance (OFR) effect in an ultracold gas of bosonic $^{88}$Sr. A
  systematic measurement of three resonances allows precise
  determinations of the OFR strength and scaling law, in agreement
  with coupled-channels theory. Resonant enhancement of the complex
  scattering length leads to thermalization mediated by
  elastic and inelastic collisions in an otherwise ideal gas. OFR
  could be used to control atomic interactions with high spatial and
  temporal resolution.
\end{abstract}

\maketitle

The ability to control the strength of atomic interactions has led to
explosive progress in the field of quantum gases for studies of few-
and many-body quantum systems. This capability is brought about by
magnetic field-induced Feshbach scattering resonances
(MFR)~\cite{chin10}, where both the magnitude and sign of low-energy
atomic interactions can be varied by coupling free particles to a
molecular state. MFR in ultracold alkali atoms have been used to
realize novel few-body quantum states and study strongly correlated
many-body systems and phase
transitions~\cite{bloch08,chin10}. %ketterle08,giorgini08,chin10}.
However, magnetic tuning has limited current experiments to relatively
slow time scales and low spatial resolution. Higher resolution could
be achieved by controlling MFR optically~\cite{bauer09}.

Scattering resonances can also arise under the influence of laser
light tuned near a photoassociation (PA) resonance~\cite{jones06}
where free atom pairs are coupled to an excited molecular
state~\cite{fedichev96,bohn99}. This Optical Feshbach Resonance (OFR)
is expected to enable new and powerful control with high spatial and
temporal resolution. OFR has been studied in thermal~\cite{fatemi00}
and degenerate~\cite{theis04,thalhammer05} gases of Rb, but it was not
found useful due to large photoassociative losses. Much narrower
optical intercombination lines are available in alkaline earth atoms
and are predicted to overcome this loss problem~\cite{ciurylo05}.
Independently, ultracold alkaline earth atoms have recently emerged to
play leading roles for quantum
metrology~\cite{akatsuka08,campbell09,poli11} where precision
measurement and many-body quantum systems are combined to study new
quantum phenomena~\cite{swallows11,gorshkov10}. Degenerate gases of
alkaline earth atoms have recently become
available~\cite{takasu03}. %,kraft09,stellmer09,escobar09}.
Due to the lack of magnetic structure in the ground state of these
atoms, the OFR effect could become an important tool for controlling
their interactions. OFR work on Yb~\cite{enomoto08,yamazaki10} has
been limited to studying the induced change in scattering phase shifts
and PA rates. Dominant PA losses are evident in all of the OFR
experiments listed above. Light-induced elastic collisions for
thermalization were not observed.

In this Letter, we study the OFR effect across multiple resonances in
a metastable molecular potential of $^{88}$Sr. The aim of this work is
to test the practical applicability of OFR for engineering atomic
interactions in the presence of loss, similar to the successful
application of a decaying MFR~\cite{naik10}. For $^{88}$Sr, OFR is
predicted~\cite{ciurylo05} to allow changes in the scattering length
by more than a factor of 100 with low losses by using large detunings
($\mathcal{O}(10^5)$ linewidths) from the least-bound vibrational
level~\cite{zelevinsky06}. We tested this proposal and find
experimentally that the existing isolated resonance
model~\cite{bohn99} only describes the experiment in the small
detuning regime. Large detunings from a molecular resonance require a
full coupled-channels description of the molecular response. Supported
by this new theory framework, we present a systematic experimental
study of the OFR-enhanced complex scattering lengths and demonstrate
OFR-induced thermalization in an ultracold gas.

\begin{figure}[htb]
  \centering
  \includegraphics[width=.9\columnwidth]{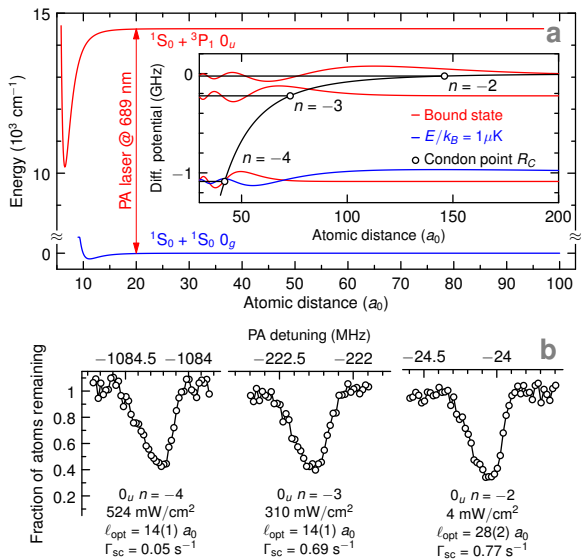}
  \caption{ (a) Ground (blue) and excited (red) molecular potentials
    of Sr$_2$. The inset shows the difference potential after
    subtracting the optical frequency. Horizontal lines indicate bound
    molecular states $n$ in the excited potential. The free particle
    (bound state) radial wave function is indicated in blue (red). (b)
    Loss spectra for $0_u~n$=-2, -3, and -4 for exposure time
    $\tau_\text{PA} = \unit{200}{ms}$ and comparable mean density. $I$
    is scaled to keep $\Gamma_\text{sc}$ sufficiently small. The
    similarity of the spectra demonstrates the universal scaling with
    $\ell_\text{opt}\propto |\langle n | E\rangle|^2 I$.}
  \label{fig:overview}
\end{figure}

Bosonic $^{88}$Sr has an $s$-wave background scattering length
$a_\text{bg} = -1.4(6) a_0$~\cite{escobar08}, where $a_0$ is the Bohr
radius. The small $|a_\text{bg}|$ makes the sample effectively
non-interacting and provides an ideal testing environment for OFR.
Figure~\ref{fig:overview}a shows the ground (\SSZ{}-\SSZ{}$~0_g$) and
lowest excited state (\SSZ{}-\TPO{}$~0_u$) molecular potentials of
Sr$_2$, which are coupled by a PA laser near the atomic transition at
$\lambda_a = \unit{689}{nm}$. The vibrational levels investigated are
labelled by their quantum number $n$, counted as negative integers
from the free particle threshold. For a given PA laser detuning from
threshold, the Franck-Condon principle localizes the atom-light
interaction in the vicinity of the Condon point~\cite{bohn99}.

When detuning the PA laser across a vibrational resonance, the
$s$-wave scattering length shows a dispersive behavior, just as for a
MFR. However, the finite lifetime of the excited molecular state leads
to loss intrinsic to OFR. This process can be described~\cite{online}
akin to decaying MFR~\cite{chin10,naik10} with a complex $s$-wave
scattering length $\alpha(k) \equiv a(k) - i b(k)$ that depends on the
relative momentum $\hbar k$ and a PA line strength factor
$\ell_\text{opt} = \frac{\lambda_a^3}{16\pi c} \frac{|\langle n | E
  \rangle|^2}{k} I$, called the optical
length~\cite{ciurylo05,ciurylo06}. Here, $c$ is the speed of light,
and $\ell_\text{opt}$ scales linearly with PA intensity $I$ and
free-bound Franck-Condon factor $|\langle n|E\rangle|^2$ per unit
collision energy $E=\hbar^2k^2/(2\mu)$ at reduced mass $\mu =
m_\text{Sr}/2$. In the isolated resonance approximation~\cite{bohn99}
the inelastic collision rate is~\cite{online}
\begin{equation}
  \label{eq:1}
  K_\text{in}(k) = \frac{4\pi\hbar}{\mu}\frac{\frac{\ell_\text{opt}\gamma_m}{\gamma}}
  {(\Delta+E/\hbar)^2/\gamma^2 +
    [1 + 2 k \frac{\ell_\text{opt}\gamma_m}{\gamma}]^2/4} \,,
\end{equation}
where $\Delta$ is the laser detuning from molecular
resonance~\cite{online}. We have accounted for extra molecular losses
with $\gamma > \gamma_m = 2\gamma_a$, where $\gamma_m$ is the
linewidth of the molecular transition and $\gamma_a =
2\pi\times\unit{7.5}{kHz}$ is the atomic linewidth. Neglecting
$a_\text{bg}$ for $^{88}$Sr gives $ K_\text{el}(k) \simeq 2 k
\frac{\ell_\text{opt}\gamma_m}{\gamma} K_\text{in}(k)$. The
elastic-to-inelastic collision ratio $K_\text{el}/K_\text{in}$ becomes
less favorable for smaller $k$.

We load $\sim$5$\times 10^4$ atoms from a magneto-optical trap
operating on the \SSZ{}-\TPO{} intercombination transition into a
crossed optical dipole trap formed by tilted horizontal (H) and
vertical (V) beams (\unit{1064}{nm}), with trap depths
$\sim$\unit{15}{\mu K} and $\sim$\unit{7}{\mu K}, respectively. The
trapped sample shows a clear kinetic energy inhomogeneity between the
H and V axes (2-\unit{2.5}{\mu K} vs. 3-\unit{4}{\mu K}), due to the
negligible $a_\text{bg}$, consistent with a thermal distribution
energy-filtered by the trap potential. Typical in-trap cloud
diameters are 45-\unit{55}{\mu m}. The PA
beam intersects the trap with a waist of \unit{41}{\mu m}~\cite{online}.

A representative survey of PA resonances in the \SSZ{}-\TPO{}$~0_u$
potential is shown in Fig.~\ref{fig:overview}b. The PA laser with
intensity $I$, adjusted to achieve similar $\ell_\text{opt}$ for all
spectra shown, interacts with the sample for $\tau_\text{PA}$.
Photon-atom scattering at rate $\Gamma_\text{sc}$ and subsequent
radiation trapping set the maximum usable $I$ for a given detuning
from the atomic line~\cite{online}. In addition to the vibrational
levels indicated in Fig.~\ref{fig:overview}a, the $n$=-1 vibrational
state exists at -\unit{0.4}{MHz} detuning from the threshold, which
leads to a PA resonance with a very large line strength
$\ell_\text{opt}/I$~\cite{zelevinsky06}. The isolated resonance theory
indicates that operating with a large $I$ at $\mathcal{O}(10^5
\gamma_a)$ detuning from the $n$=-1 state should allow modifications
to $a(k)$ of $\mathcal{O}(100 a_0)$~\cite{ciurylo05}. This prediction
relied on extrapolating the large line strength of the $n$=-1 state
across multiple intermediate PA resonances. However, with $I$ up to
\unit{1}{kW/cm^2} and detunings up to -\unit{1.5}{GHz}, we did not
observe any effects due to elastic collisions.

The discrepancy between theory and experiment stimulated a
coupled-channels treatment of an atomic collision in a radiation field
that properly switches between the short range molecular states and
two field-dressed separated
atoms~\cite{online,julienne86,napolitano97}. In the coupled-channels
theory, the two coupled excited potentials ($0_u$, $1_u$) have the
form of Ref.~\cite{zelevinsky06}, with an added imaginary term
$-i\hbar\gamma_m/2$. The ground state potential uses the dispersion
coefficients of Ref.~\cite{porsev06}, has a scattering length of
$-1.4~a_0$, and reproduces the bound state data of
Ref.~\cite{escobar08} to better than 0.4\%. Coupled-channels
calculations do not assume isolated resonances, and all $0_u$ and
$1_u$ molecular eigenstates emerge from the calculation as
interfering, decaying scattering resonances~\cite{chin10}.

\begin{figure}[htb]
  \centering
  \includegraphics[width=\columnwidth]{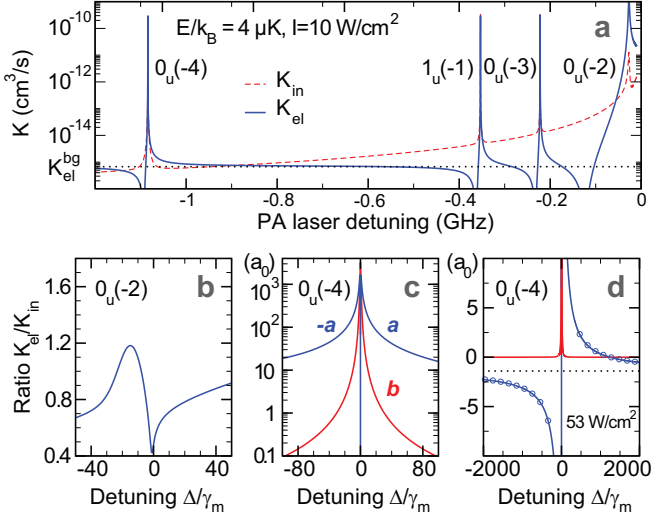}
  \caption{ (a) Coupled-channels calculations of $K_\text{el}$ and
    $K_\text{in}$ at $E/k_B=\unit{4}{\mu K}$ and
    $I=\unit{10}{W/cm^2}$, versus PA laser detuning from atomic
    resonance. Each resonance peak is labeled by its electronic
    symmetry $0_u$ or $1_u$ and $n$. Between resonances, $K_\text{in}$
    is only approximate. (b) Ratio of thermally-averaged rate
    constants at \unit{2}{\mu K} for $\Delta/\gamma_m$ near $0_u(-2)$.
    $I = \unit{44}{mW/cm^2}$ gives the same $\ell_\text{opt}=360 a_0$
    as for the conditions in Fig.~\ref{fig:elastic}b. (c) Zero energy
    limit of $a(k)$ and $b(k)$ for detuning near the $0_u(-4)$
    feature. (d) Same calculation as in (c) at large detuning. Here, the
    isolated resonance results (solid lines) agree with the
    coupled-channels theory (circles).}
\label{fig:theory}
\end{figure}

Figures~\ref{fig:theory}c and d show that the coupled-channels model
reproduces the isolated resonance expressions~\cite{bohn99,chin10} for
$\alpha(k)$ and the rate constants as long as $\Delta$ is small
compared to the spacing between molecular levels. However, the
coupled-channels $K_\text{el}$ returns to its background value
$K_\text{el}^\text{bg}$ in between resonances regardless of their
relative strengths. The dotted line indicates $K_\text{el}^\text{bg}$
($a_\text{bg}$) at $E/k_B = \unit{4}{\mu K}$ in Fig.~\ref{fig:theory}a
(Fig.~\ref{fig:theory}d). These calculations show that each molecular
line behaves as an isolated resonance near its line center. For
detunings comparable to the molecular level spacing, the isolated
resonance expressions cannot be used.

At intermediate detunings, $|\Delta| \gg \gamma(1 + 2 k
\ell_\text{opt} \frac{\gamma_m}{\gamma})$, $\alpha(k)$ can be written
in the standard form for MFR~\cite{online},
\begin{equation}
  \label{eq:6}
  \lim_{k\to 0}\alpha(k) = a_\mathrm{bg} \left ( 1 - \frac{w}{\Delta} + \frac{i}{2}
    \, \frac{w\gamma}{\Delta^2} \right ),
\end{equation}
where $w \equiv -\ell_\mathrm{opt} \gamma_m/a_\text{bg}$. To obtain a
meaningful change in scattering length,
$\ell_\text{opt}\gamma_m/\Delta$ needs to be sufficiently large, and
the imaginary part $b=\frac{1}{2} \ell_\mathrm{opt} \gamma_m
\gamma/\Delta^2$ needs to be sufficiently small. Since
$K_\text{in}\simeq (\unit{2\times 10^{-12}}{cm^3/s})~(b/a_0)$, for a
density of $\rho=\unit{10^{12}}{cm^{-3}}$ and $b=0.1 a_0$,
$K_\text{in}\rho = \Gamma_\text{sc}$ for $I=\unit{53}{W/cm^2}$ assumed
for Figs.~\ref{fig:theory}c,d. Thus, the calculations predict that
changes in the scattering length of order 10$~a_0 \gg | a_\text{bg} |
$ should be possible with $\mathcal{O}(100~\gamma_m)$ detunings on
timescales of \unit{200}{ms}.

To investigate the utility of OFR, we systematically characterized
three different resonances and determined their universal scaling. Because
$K_\text{el}/K_\text{in}\propto k \ell_\text{opt}$, inelastic
collisions dominate the dynamics of the sample for small
$\ell_\text{opt} \ll (2\langle k\rangle)^{-1} =
\frac{\hbar}{2}\sqrt{\pi/(8\mu k_B T)}$, where the angled brackets
indicate a $k$-average at temperature $T$, and $k_B$ is the Boltzmann
constant. In this regime, the result of scanning the PA laser across
resonance is a loss feature that shows no dependence on elastic
collision processes.

\begin{figure}[htb]
  \centering
  \includegraphics[width=\columnwidth]{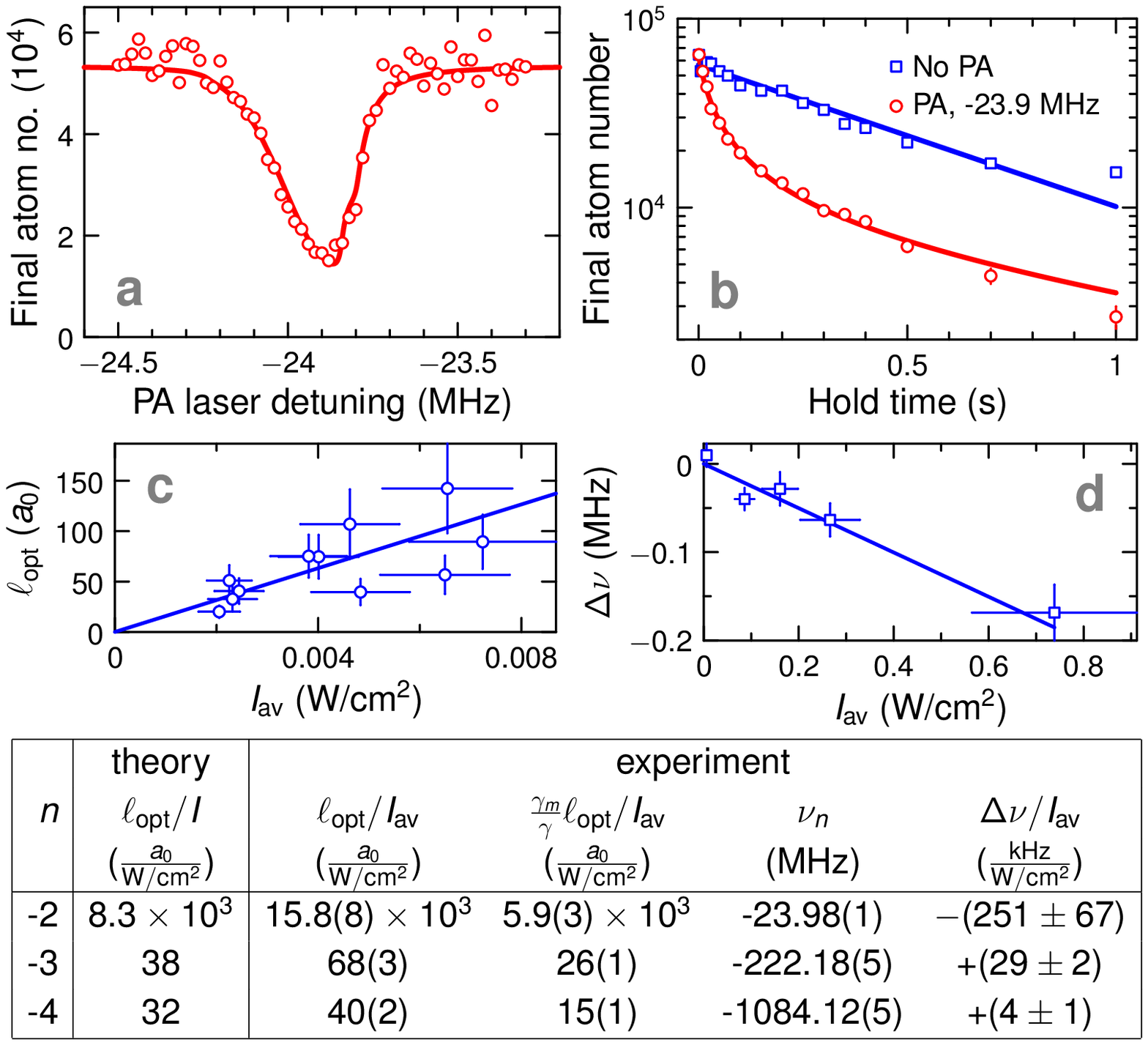}
  \caption{(a) Typical PA loss feature for $n$=-2 in the low intensity
    regime at $I_\text{av} = \unit{7}{mW/cm^2}$, with
    density-profile-averaged PA intensity $I_\text{av}$~\cite{online}.
    (b) Time evolution of the trapped sample with (circles) and
    without PA light (squares). The two-body loss curve with PA is fit
    with a thermally averaged model (solid curve). (c) Linear increase
    of $\ell_\text{opt}$ with $I_\text{av}$ for $\ell_\text{opt}\ll
    1/(2\langle k\rangle)$. (d) Molecular line center shift
    $\Delta\nu$ for large $I_\text{av}$ and decreased
    $\tau_\text{PA}$. For each $n$, OFR parameters from the
    coupled-channels calculation and the experiment are summarized in
    the table at the bottom. Here, $\nu_n$ is the zero-intensity
    molecular line center with respect to \SSZ{}-\TPO{}, and
    $\Delta\nu/I_\text{av}$ characterizes the molecular ac Stark
    shift~\cite{ciurylo06}.}
  \label{fig:lopt}
\end{figure}

A typical PA loss feature for small $\ell_\text{opt}$ is shown in
Fig.~\ref{fig:lopt}a, where the final atom number after application of
PA light is shown with respect to PA detuning from \SSZ{}-\TPO{}. The
per-axis kinetic energies~\cite{online} for this scan correspond to a
horizontal (vertical) temperature $T_H (T_V) = \unit{2}{\mu
  K}~(\unit{3}{\mu K})$, resulting in the typical thermal tail towards
the red side of the resonance~\cite{ciurylo04}. The solid line is a
result of solving~\cite{burke99}
\begin{equation}
  \label{eq:4}
  \dot{\rho}(\vec{x},t) = - 2\bar{K}_\text{in}(\vec{x})\frac{\rho^2(\vec{x},t)}{2}
  - \frac{\rho(\vec{x},t)}{\tau_\text{bg}},
\end{equation}
with single particle density $\rho$, thermally-averaged inelastic
collision rate $\bar{K}_\text{in}(\vec{x}) \equiv \langle
K_\text{in}(k,\Delta,\vec{x}) \rangle$~\cite{online}, and
vacuum-limited trap lifetime $\tau_\text{bg}\simeq\unit{1.3}{s}$.
Equation~\ref{eq:4} is solved at each $\vec{x}$ and we integrate the
density to obtain the number of atoms at the end of the PA pulse.
We then use Eq.~\ref{eq:4} to fit to the experimental
data~\cite{online} and extract $\ell_\text{opt}$ and the position of
the line center. Figure~\ref{fig:lopt}b shows the time dependence of
the same photoassociative loss process. Two-body loss is evident under
resonant PA light.

From the experimental data we extract two independent quantities:
$\ell_\text{opt}\gamma_m$ and an increased molecular loss rate $\gamma
\simeq 2.7 \gamma_m$. We have ruled out magnetic field or PA laser
noise as a source of broadening. Instead, we conclude that this extra
broadening is related to a faster molecular decay rate, which is
consistent with our earlier measurements~\cite{zelevinsky06} and Rb
results~\cite{theis04}.

The measurements were performed for a range of $\ell_\text{opt}$ by
adjusting $I_\text{av}$. Multiple molecular resonances were measured
and results for $n$=-2 are shown in Fig.~\ref{fig:lopt}c. The optical
length data is fit with a linear function and the results are
summarized in the table at the bottom. The fit coefficient
$\ell_\text{opt}/I_\text{av}$ is given by the free-bound Franck-Condon
factor and decreases drastically with decreasing $n$.
Figure~\ref{fig:lopt}d exemplifies similar measurements done to
determine the line shift $\Delta\nu$ with $I_\text{av}$. Linear shift
coefficients $\Delta\nu/I_\text{av}$ and zero intensity line positions
$\nu_n$ with respect to the atomic transition are also shown in the
table. The sign and magnitude of $\Delta\nu/I_\text{av}$ are
consistent with the predictions in Ref.~\cite{ciurylo06}.

\begin{figure}[htb]
  \centering
  \includegraphics[width=.9\columnwidth]{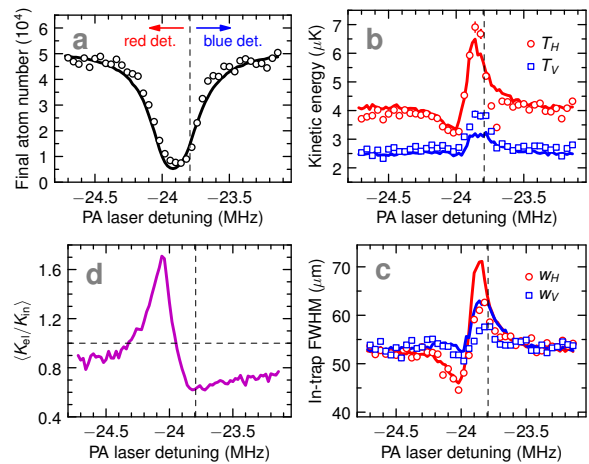}
  \caption{ Elastic contribution to the scattering
    cross section for $n$=-2 at $I_\text{av} = \unit{22}{mW/cm^2}$
    (open circles) and results of a Monte-Carlo simulation (solid
    lines) using Eq.~\ref{eq:1} in a crossed dipole trap for
    $\ell_\text{opt}\gamma_m/\gamma = 140 a_0$. (a) Atom loss as a
    function of PA laser detuning from the atomic \SSZ{}-\TPO{}
    resonance. In panels b and c, blue (red) data points and solid
    lines indicate the corresponding quantities for the vertical
    (horizontal) trap axis. (b) Change in kinetic energy derived from
    time-of-flight images, and (c) potential energy change
    corresponding to varying in-trap density profile. (d) The
    resultant ratio of elastic and inelastic collisions per particle,
    averaged over $\tau_\text{PA}$.}
  \label{fig:elastic}
\end{figure}

At larger optical lengths ($\ell_\text{opt}\gamma_m/\gamma\sim 100
a_0$), elastic collisions start to influence the dynamics of the
system. We show the atom loss with respect to PA laser detuning for
$n$=-2 in Fig.~\ref{fig:elastic}a. Both in-trap size and kinetic
energy are measured by absorption imaging~\cite{online}. Far detuned
from the resonance, the gas is almost ideal, as shown by the
persistent kinetic energy inhomogeneity along H and V in
Fig.~\ref{fig:elastic}b. On resonance (vertical dashed line),
inelastic collisions dominate and cause heating. For red detuning from
the molecular resonance, the temperatures approach each other by
cross-thermalization~\cite{goldwin05}.

The measured cloud widths $w_H$ and $w_V$ confirm that the potential
energy follows the kinetic energy (Fig.~\ref{fig:elastic}c) since
particles oscillate in the trap many times between collisions. Similar
measurements were performed for $n$=-3 and $n$=-4, and we find that
the same dispersive behavior in temperatures and widths appears around
$2 \langle k \rangle \ell_\text{opt}\gamma_m/\gamma \sim 30\%$ at
$\tau_\text{PA} = \unit{200}{ms}$. The data can be understood by a
simple picture of competition between $K_\text{el}$ and $K_\text{in}$,
which average differently with $k$ in a thermal sample [see
Eq.~(\ref{eq:1})], and thus peak at different values of $\Delta$.
Elastic collisions cause cross-dimensional thermalization and tend to
equalize $T_H$ and $T_V$. Since inelastic collisions predominantly
remove cold atoms from the densest part of the cloud, the resulting
loss increases the average system energy via anti-evaporation.

This behavior is confirmed by a Monte-Carlo simulation, where
$55\times10^3$ particles are simulated and each particle undergoes
elastic and inelastic collisions with an initial phase-space distribution
matched to the experimental conditions~\cite{online}. The
solid lines overlaid on the experimental data in
Fig.~\ref{fig:elastic} are the simulation results. An average ratio of
elastic to inelastic collisions per particle from the simulation is
shown in Fig.~\ref{fig:elastic}d. The dispersive shapes are also
predicted by the coupled-channels model (see Fig.~\ref{fig:theory}b)
and their shape is sensitive to $\gamma$. Combined with the low
$\ell_\text{opt}$ data in Fig.~\ref{fig:lopt}, the entire simulation
reproduces the experimental data only when $\gamma =
2\pi\times\unit{40(5)}{kHz}$ without other free parameters.

We conclude that the isolated resonance approximation universally
describes OFR in the vicinity of each resonance. The coupled-channels
calculation includes all interference effects between resonances, and
differs from the isolated resonance approximation at large detuning
between resonances. Our experiment contradicts previous predictions
based on extrapolations of an isolated resonance to large
detunings~\cite{ciurylo05,zelevinsky06}. We have validated the linear
line strength scaling and linear resonance shift with $I$ and have
observed a clear modification of both $a(k)$ and $b(k)$. For the
values of $\ell_\text{opt}\gamma_m/\gamma$ achieved here, inelastic
losses still contribute significantly and $\langle
K_\text{el}/K_\text{in} \rangle$ becomes even less favorable with
decreasing $T$. However, the OFR effect can modify interactions in a
degenerate gas of alkaline earth atoms and the desired change of
$a(k)$ is achieved at the smallest $\Delta/\gamma$ constrained by both
molecular and atomic loss processes over a given experimental
timescale~\cite{online}.

We thank C.~Greene, P.~Zoller, and G.~Campbell for discussions and
contributions. J.~R.~W. is an NRC Fellow. Our work is funded by DARPA
OLE, NIST, \& NSF.

\appendix
\section{SUPPLEMENTARY MATERIAL}

\section{Scattering matrix, cross sections, and collision rates}
\label{sec:s-matrix-scattering}

The collision of two identical bosons in the low energy near-threshold
limit can be described in a scattering matrix treatment by a single
$s$-wave scattering matrix element, $S(k) = e^{2 i \eta(k)}$, where
$\hbar k$ is the relative momentum of the collision pair and $\eta(k)$
is the scattering phase shift due to interactions. When there is loss
of scattering flux from the entrance channel, as is the case of a
decaying optical Feshbach resonance, it is convenient to {\em define}
a complex energy-dependent scattering length $\alpha(k)$, related to
the complex phase $\eta(k)$ and $S$-matrix element
by~\cite{om:hutson07,om:chin10}
\begin{equation}
 \label{complexA}
  \alpha(k) \equiv a(k) - i b(k) \equiv
  -\frac{\tan{\eta(k)}}{k} = \frac{1}{ik}
  \frac{1-S(k)}{1+S(k)}.
\end{equation}
This reduces to the usual definition of complex scattering length as $k \to 0$.
The respective elastic and inelastic $s$-wave collision cross sections
are~\cite{om:chin10,om:burke99}
\begin{equation}
  \label{om:eq:9}
  \begin{split}
    \sigma_\text{el}(k) &= \frac{2\pi}{k^2} |1-S(k)|^2 =
   8\pi |\alpha(k)|^2 f^2(k), \\
    \sigma_\text{in}(k) &= \frac{2\pi}{k^2} (1-|S(k)|^2)  =
    \frac{8\pi}{k} b(k)f(k), \\
  \end{split}
\end{equation}
where $f(k)=[1+k^2|\alpha(k)|^2 + 2kb(k)]^{-1} \to 1$ as $k \to 0$ and
$\mu$ is the reduced mass of the pair. The second relations in
Eq.~\ref{om:eq:9} follow from the definition in Eq.~\ref{complexA}. The
corresponding collision rate coefficients are
\begin{equation}
  \label{om:eq:10}
  \begin{split}
    K_\text{el}(k) &= \frac{\hbar k}{\mu} \sigma_\text{el}(k), \\
    K_\text{in}(k) &= \frac{\hbar k}{\mu} \sigma_\text{in}(k). \\
  \end{split}
\end{equation}

When the entrance channel is coupled by a single frequency laser to a
single excited molecular bound state of the pair, the isolated
resonance approximation for the the field-dressed $s$-wave scattering
matrix element can be written as~\cite{om:bohn99,om:ciurylo06},
\begin{equation}
\label{om:eq:5}
  S(k) \simeq e^{2i\eta_\text{bg}(k)}\left( 1- i \frac{\Gamma_s(k)}{E/\hbar + \Delta
    + i (\gamma + \Gamma_s(k))/2}\right).
\end{equation}
Here $E = \hbar^2 k^2 / 2 \mu$ is the collision energy with $\mu =
m_\text{Sr}/2$. The laser detuning $\Delta$ from the PA resonance
includes the light shift~\cite{om:ciurylo06} to simplify the presentation
(see Eq.~\ref{om:eq:2} for a full expression).

The stimulated width $\Gamma_s(k) \equiv 2 k \ell_\text{opt}(k)
\gamma_m$ is the induced coupling between the free particle state
$|E\rangle$ and the excited molecular state $|n\rangle$. Here
$\gamma_m = 2 \gamma_a$ is the linewidth of the molecular transition
and $\gamma_a = 2\pi\times\unit{7.5}{kHz}$ is the atomic linewidth. We
have allowed for extra molecular losses by letting $\gamma >
\gamma_m$. The optical length
\begin{equation}
\label{om:eq:6}
  \ell_\text{opt} = \frac{\lambda_a^3}{16\pi c}
  \frac{|\langle n | E \rangle|^2}{k} I,
\end{equation}
where $c$ is the speed of light, is a PA resonance-dependent line
strength parameter~\cite{om:ciurylo05,om:ciurylo06}. Under the Wigner
threshold law, the free-bound Franck-Condon factor per unit collision
energy $|\langle n|E\rangle|^2 \propto k$, and we checked numerically
that it is a very good approximation to take $\ell_\text{opt}$ to be
independent of collision energy for temperatures $<\unit{10}{\mu K}$.
The optical length is also independent of the atomic oscillator
strength, and it scales linearly with PA intensity $I$.

Combining Eq.~\ref{om:eq:5} with Eqs.~\ref{complexA}-\ref{om:eq:10} gives
the near-threshold isolated resonance approximation to the OFR
inelastic loss rate coefficient:
\begin{equation}
\label{om:eq:7}
    K_\text{in}(k) = \frac{4\pi\hbar}{\mu}\frac{\frac{\ell_\text{opt}\gamma_m}{\gamma}}
    {(\Delta+E/\hbar)^2/\gamma^2 +
    [1 + 2 k \frac{\ell_\text{opt}\gamma_m}{\gamma}]^2/4} \,.
\end{equation}
The background $s$-wave scattering length $a_\text{bg}$ is defined in
the absence of a resonance as
\begin{equation}
  \label{om:eq:8}
  a_\text{bg}=-\lim_{k\to 0}\frac{\tan{\eta_\text{bg}(k)}}{k}.
\end{equation}
If we neglect $a_\text{bg}$ for $^{88}$Sr, we find a simple expression
for the OFR-induced elastic collision rate coefficient for an ideal
ultracold gas
\begin{equation}
  \label{om:eq:11}
  K_\text{el}(k) \simeq 2 k \frac{\ell_\text{opt}\gamma_m}{\gamma} K_\text{in}(k).
\end{equation}

Power broadening is included via $\Gamma_s(k)$ in the denominator of
Eq.~\ref{om:eq:5}, which corresponds to $2k
\frac{\ell_\text{opt}\gamma_m}{\gamma}$ in the denominator of
Eq.~\ref{om:eq:7}. From the expressions for elastic and inelastic
collision rates, we see that the relevant strength parameter in the
presence of extra molecular loss is the rescaled optical length
$\ell_\text{opt}\frac{\gamma_m}{\gamma}$. Elastic collisions and power
broadening only become important for the dynamics when
\begin{equation}
  \label{om:eq:15}
  2 k \frac{\ell_\text{opt} \gamma_m}{\gamma} \sim 1.
\end{equation}

\section{Coupled Channels Calculations}

The simplest way to do a calculation of the elastic and inelastic
collision properties of an optical Feshbach resonance is to set up a
coupled channels model for the collision of field-dressed states in a
laser field of fixed frequency~\cite{om:julienne86,om:bohn99}. We use the
simplest three-channel model that is sufficiently complete to
represent nonperturbatively the optically induced $S(k)$ for an
interfering spectrum of excited state resonances. One channel
represents the ground state with potential $V_g(R)$ and two channels
represent the two excited states $0_u$ and $1_u$ that correlate with
the separated atom limit $^1$S$_0$ $+$ $^3$P$_1$ with respective
complex potentials $V_{0u}$ and $V_{1u}$. These potentials are shifted
to include the asymptotic detuning of the laser frequency from atomic
resonance. It is necessary to include {\em both} $0_u$ and $1_u$
excited states to correctly calculate the complex scattering length
near an isolated resonance of either state. This is because the
molecular $J=1$ excited states (J= total angular momentum quantum
number) must become a mixture of $s$ and $d$-waves at long range to
properly go to separated atoms with $^3$P$_1$ quantized in a space
frame instead of a molecular body frame~\cite{om:julienne86}; otherwise,
spurious $1/R^3$ resonant dipole terms would get mixed into the ground
state potential and give invalid threshold scattering lengths.

The coupled channel potential matrix $\bm{V}(R)$ is found using the
asymptotic representation of the two excited $J=1$ molecular channels
in terms of a pair of excited state $s$- and $d$-waves for the $e$
parity block in Table 1 of Reference~\cite{om:julienne86}:
\begin{equation}
\label{Vmatrix}
  \bm{V} = \left( \begin{array}{ccc}
      V_g & V_\mathrm{opt} & 0  \\
      V_\mathrm{opt} &  \frac{1}{3}(V_{0u}+2V_{1u}) & \frac{\sqrt{2}}{3}(V_{1u}-V_{0u}) \\
      0 &  \frac{\sqrt{2}}{3}(V_{1u}-V_{0u}) &  \frac{1}{3}(2V_{0u}+V_{1u}+6B)
    \end{array} \right ),
\end{equation}
where $6B=6\hbar^2/(2\mu R^2)$ is the $d$-wave centrifugal energy.
Here the row labels, in order from top to bottom, represent the
$|j\ell JM\rangle$ $=$ $|0000\rangle$ ground state and $|1010\rangle$
and $|1210\rangle$ excited state channels of Ref.~\cite{om:julienne86},
where $j$ is the atomic angular momentum for the collision partner to
the $^1$S$_0$ atom, $\ell$ is partial wave, and $JM$ represent the
total angular momentum and its projection for the pair of interacting
atoms. Since the excited bound states are relatively short range, we
use the nonretarded optical coupling $V_\mathrm{opt}=(2\pi I/c)^{1/2}
d_m$, where $d_m = \sqrt{2} d_A$ and the atomic transition dipole
$d_A=0.086816$ atomic units corresponds to an atomic lifetime of 21.46
$\mu$s. A complete theory would use the retarded optical coupling. The
zero of energy for ground state collisions is taken to be the lowest
eigenvalue of this matrix to account for the energy shift of the field
dressed molecule. The numerical $S$-matrix is calculated using
standard coupled channels methods to calculate the wave function for
the coupling matrix in Eq.~\ref{Vmatrix}.

While Reference~\cite{om:bohn99} simulates the spontaneous radiative
decay of the excited state by introducing artificial channels, here we
include a complex term $-i\hbar\gamma/2$ in each of these excited
state potentials to simulate this decay, where in our numerical
calculations we take $\gamma=\gamma_m$. This leads to a non-unitary
ground state $S$-matrix element such that $0 \le 1-|S(k)|^2 \le 1 $
gives probability of atom loss during a collision. In order to avoid
problems with spurious asymptotic decay of the field dressed states, a
tapered cutoff function is introduced so that decay is only turned on
inside some characteristic distance, which we take to be 500 a$_0$. We
have verified that $a(k)$ is independent of this choice over the full
range of detunings, even between resonances, and $b(k)$ is independent
of this choice out to many linewidths away from an isolated resonance.
However, in the far wings of a resonance between isolated resonances,
$b(k)$ will depend on this choice, although molecular loss tends to be
small in these regions. This dependence on cutoff is because our
non-unitary theory does not distinguish between atomic and molecular
light scattering when the atoms are far apart; the cutoff selects
which distance inside of which an excitation is considered to be a
molecular loss process instead of an atomic one. Thus the cutoff is
taken to be on the order of but larger than the outer turning point of
the excited molecular bound states.

Finally, we ignore coupling of the excited $J=1$ molecular levels back
to $d$-waves in the ground state. This means that our model only gets
the $s$-wave but not the $d$-wave contributions to the excited state
light shifts~\cite{om:ciurylo06}. It would be straightforward to add an
extra ground state channel to the numerical calculation to account for
this.

While there are a number of ways our simple model can be improved, it
gives a practical way to calculate the needed complex scattering
length for characterizing the intercombination line OFR spectrum of
alkaline earth atoms. The model gives two key results. First, it
should correctly describe the variation of the real part of the
scattering length as laser frequency is tuned across multiple
resonances. Second, the isolated resonance approximation should
correctly describe the molecular loss rate out to order of hundreds of
line widths away from individual isolated resonances.

\section{Experimental Setup}
\label{sec:experimental-setup}

\begin{figure}[htb]
  \centering
  \includegraphics[width=\columnwidth]{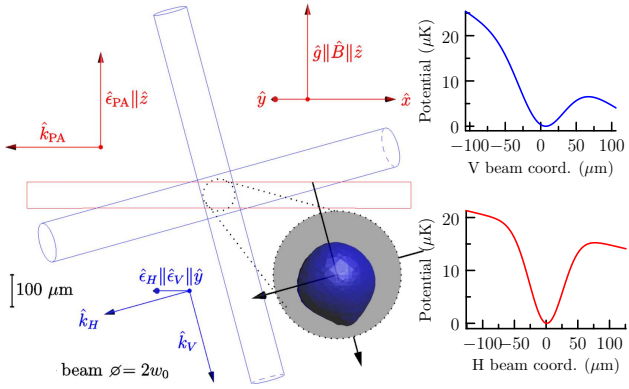}
  \caption{ Geometry of the experiment in the absorption image plane.
    The directions $\hat{x}$, $\hat{y}$, and $\hat{z}$ define the lab
    frame, where both gravity $\hat{g}$ and bias magnetic field
    $\hat{B}$ are parallel to $\hat{z}$. Symbols $\hat{k}$ are beam
    directions, and $\hat{\epsilon}$ are beam polarization vectors,
    where subscripts $H$, $V$, and $\mathrm{PA}$ indicate horizontal,
    vertical, and PA beams. $H$ and $V$ Gaussian beam profiles are
    shown in blue outline, PA Gaussian beam profile in red outline. }
  \label{om:fig:setup}
\end{figure}

Details of the experimental setup can be found in Ref.~\cite{om:blatt11}.
Atoms are loaded into an optical dipole trap from a magneto-optical
trap operating on the \SSZ{}-\TPO{} intercombination transition. A
horizontal (H) and a vertical (V) beam intercept in the
$\hat{x}-\hat{z}$ plane to form a crossed dipole trap at the origin
shown in Fig.~\ref{om:fig:setup} (drawing is to scale). Both beams are
derived from the same \unit{1064}{nm} laser and are linearly polarized
along the $\hat{y}$ axis with $1/e^2$ waists \unit{63}{\mu m} and
\unit{53}{\mu m}, respectively. The optical beams point in the
directions
\begin{equation}
  \label{om:eq:3}
  \begin{aligned}
  \hat{\vec{k}}_H &= -\cos\theta_H \hat{\vec{x}} - \sin\theta_H \hat{\vec{z}},\\
  \hat{\vec{k}}_V &= -\cos\theta_V \hat{\vec{z}} + \sin\theta_V
  \hat{\vec{x}},\\
  \hat{\vec{k}}_\text{PA} & = -\cos\theta_\text{PA} \hat{\vec{x}} +
  \sin\theta_\text{PA} \hat{\vec{y}},\\
  \end{aligned}
\end{equation}
with $\theta_H = 16.0^\circ$, $\theta_V = 14.4^\circ$, and
$\theta_\text{PA} = \pi/8$. The beams are linearly polarized along
$\hat{\vec{\epsilon}}_H = \hat{\vec{\epsilon}}_V = \hat{\vec{y}}$ and
$\hat{\vec{\epsilon}}_\text{PA} = \hat{\vec{z}}$. The image plane is
spanned by
$\cos\frac{\pi}{8}\hat{\vec{x}}-\sin\frac{\pi}{8}\hat{\vec{y}}$ and
$\hat{\vec{z}}$. A bias magnetic field of $B_z
\simeq \unit{100}{mGauss}$ defines the atomic quantization axis.

The focal condition is not critical to model the potential
sufficiently well because of the large Rayleigh ranges
($>\unit{8}{mm}$), and the Gaussian beam isosurfaces at the beam waist
are almost cylindrical (blue outlines). Thus, the dipole trap
potential due to each beam is modeled as
\begin{equation}
\label{om:eq:12}
  U_i(\vec{x}) \equiv - U_T^i \exp\left\{-\frac{2}{w_i^2}
    \left([\hat{\vec{\epsilon}}_i\cdot\vec{x}]^2 + [(\hat{\vec{k}}_i \times
    \hat{\vec{\epsilon}}_i)\cdot\vec{x}]^2\right)
  \right\},
\end{equation}
with individual Gaussian beam trap depth~\cite{om:grimm99}
\begin{equation}
\label{om:eq:14}
  U_T^i = \frac{P_i}{\pi c \epsilon_0 w_i^2} \text{Re}~\alpha_{\SSZ{}}(\unit{1064}{nm}),
\end{equation}
and total laser power $P_i$. The real part of the dynamic \SSZ{}
polarizability at \unit{1064}{nm} is
$\text{Re}~\alpha_{\SSZ}(\unit{1064}{nm}) \simeq
\unit{239}{a.u.}$~\cite{om:boyd07c}, with atomic unit of polarizability
$4\pi\epsilon_0 a_0^3$.

The full model potential includes the gravitational acceleration $g$
pointing along $-\hat{z}$ and is given by
\begin{equation}
\label{om:eq:13}
  U(\vec{x}) = U_H(\vec{x}) + U_V(\vec{x}) + m_\text{Sr} g z.
\end{equation}
Gravity sets the trap depths of $\sim$\unit{7}{\mu K} and
$\sim$\unit{15}{\mu K} along V and H. The two graphs on the right hand
side of Fig.~\ref{om:fig:setup} show cuts through the model potential,
which has been adjusted to match the measured trapping frequencies. An
isosurface (dark blue) of the model potential at \unit{7}{\mu K} is
shown in the zoomed-out portion.

Typical kinetic energies are 2-\unit{4}{\mu K}, and in-trap Full
Widths at Half Maximum (FWHM) are 45-\unit{55}{\mu m}. The PA beam
(red outline) propagates along the horizontal image axis in the
$\hat{x}-\hat{y}$ plane with a waist $w_\text{PA} = \unit{41}{\mu m}$
and is linearly polarized along $\hat{z}$. Although the PA beam
diameter $2 \sqrt{2\ln 2} \times w_\text{PA} \simeq \unit{97}{\mu m}$
is larger than the typical cloud FWHM, we use a density-averaged
intensity $I_\text{av} = \int d^3 x \rho(\vec{x}) I(\vec{x}) /\int d^3
x \rho(\vec{x})$ to characterize the PA intensity interacting with the
atoms. Typical values are $I_\text{av} \simeq (0.6-0.7) \times
I_\text{pk}$, where $I_\text{pk} = \frac{2 P}{\pi w_\text{PA}^2}$ is
the Gaussian beam peak intensity for total beam power $P$.

In addition to scattering from the atomic transition with rate
$\Gamma_\text{sc}$, the PA beam also adds to the optical trap
slightly. For large intensities and small $w_\text{PA}$ and especially
in a standing wave configuration, this effect can become important. In
this work, the additional trap depth introduced by the PA beam is
typically $< \unit{0.1}{\mu K}$.

\section{Imaging Procedure and Analysis}

We use a two-component long-working-distance microscope with
\unit{3}{\mu m} resolution at \unit{532}{nm}. The imaging lens is
placed outside the vacuum chamber at a distance of \unit{150}{mm} from
the atomic cloud. The second part of the telescope is an eyepiece
mounted to a CCD camera with $1024\times 1024$ square pixels of
\unit{13}{\mu m} width. The eyepiece includes an interference filter
at \unit{461}{nm} (intensity transmission 0.45 within \unit{10}{nm}
bandwidth, ND 5.0 otherwise). Time of flight measurements were used to
determine an imaging magnification of 6.5, corresponding to
\unit{2.0}{\mu m} per pixel. We tested the imaging system resolution
in a separate test setup that included the effect of an
anti-reflection (AR) coated vacuum viewport, same as the one used on
our vacuum chamber. Using incoherent white light filtered by the blue
interference filter with transmissive test targets, resolutions of
$\sim\unit{4}{\mu m}$ were measured. Absorption imaging with coherent
light causes etaloning between the CCD chip and the glass plate that
covers it. To reduce fringing, the glass plate surface is wedged at
$\sim 1^\circ$ and AR coated for the imaging wavelength.

Without the elastic contribution of the OFR effect, the $^{88}$Sr
sample does not thermalize on experimental timescales due to its small
$a_\text{bg}$, and it is important to obtain information about both
the potential and kinetic energies of the gas. We interleave
experimental runs between in-situ imaging of the sample and imaging
after the atoms were dropped for \unit{1.5}{ms} by switching off the
optical dipole trap. The first image gives information about the
spatial density of the atomic cloud, and the second image measures the
kinetic energy by time-of-flight expansion. We typically repeat this
interleaved sequence 3-5 times and average the resulting in-situ and
time-of-flight absorption images before performing data analysis. We
also calculate the pixel-by-pixel studentized standard deviations
which are used as weights in fitting the mean image and also enter the
fit parameter error estimates. Both the mean and standard-deviation images
are then rotated (with bilinear interpolation) by $14^\circ$ about the
center of the atom cloud to transform the images into the eigenframe of
the trap. The pictures are then integrated along the horizontal
(vertical) axis and are fit to extract $w_H$ ($w_V$) with
one-dimensional Gaussian distributions including a background offset
and linear slope to account for residual fringing and CCD readout
noise in the absorption image.

A common problem in imaging atomic clouds with large optical depth is
light that does not interact with atoms hitting the camera in the same
position. Common causes are forward scattering of photons into the
imaging path or mixed probe polarization. Probing the \SSZ{}-\SPO{}
transition is insensitive to probe polarization even under a small
bias magnetic field. To identify the possibility of a limiting optical
depth OD$_\text{ceiling}$, a series of images were taken with
increasing atomic density. We compared the peak optical depth in-situ
versus the pixel-summed optical depth either of an in-situ or a
corresponding time-of-flight image, and we estimate a conservative
limit OD$_\text{ceiling} \ge 3.5$. Typical in-situ optical depths in
this paper were below 2, so that we can neglect the effect.

For probe intensities larger than $0.1 I_\text{sat}$, with saturation
intensity $I_\text{sat} \simeq \unit{40}{mW/cm^2}$, saturation of the
imaging transition becomes important. To account for the saturated
absorption, we correct the full expression for the OD to the linear
Beer-Lambert regime~\cite{om:allen87,om:blatt11}. For each
picture the saturation correction is applied on a pixel-by-pixel
basis.

While the shadow image is forming and the atoms are scattering light,
their momentum undergoes a random walk due to the spontaneous emission
of blue photons. By examining the shadow image versus a reference
image without atoms, we extract the number of photons that the atoms
removed from the probe pulse. By varying the probe time, we find the
cloud expansion as a function of the number of photons scattered and
apply a corresponding correction to the cloud widths from both in-situ
images and TOF images~\cite{om:blatt11}.

We extract the integrated OD of the mean image to obtain the atom
number. Corrections for probe heating are then applied to the fitted
widths. The in-situ image widths are used to calculate the mean atomic
density. Finally, we calculate a time-of-flight temperature corrected
for finite size effects by the in-situ widths.

\section{Loss Spectra ($2 k \ell_\text{opt}\gamma_m/\gamma \ll 1$)}
\label{sec:loss-spectra}

To extract $\ell_\text{opt}\gamma_m$ from the loss spectra, data was
fit to an approximate expression for the integral of the atomic
density, where the density after the PA pulse was calculated via the
differential equation $\dot{\rho} = - \bar{K}_\text{in} \rho^{2} -
\rho / \tau_\text{bg}$~\cite{om:zelevinsky06}. The initial density
distribution was modelled as Gaussian with $w_H$ and $w_V$. The width
into the plane of the images was extrapolated from a Monte-Carlo
simulation of the trap model potential. The fitting function is
\begin{equation}
  N = \frac{2}{ \sqrt{\pi} } \int_{0}^{\infty} \!\!\! du~
  \frac{ \sqrt{u} \, \, e^{-u} }{1 +
    \bar{K}_\text{in}(\Delta,u,\ell_\text{opt}\gamma_m,T) \,
    \tau_\text{eff} \, \rho_{0} e^{-u}},
\end{equation}
with normalized dimensionless trap length scale $u$. This expression
describes the number signal normalized to the measured atom number
after the PA pulse of duration $\tau_\text{PA}$ and when the detuning
from molecular resonance is large. The quantity $\tau_\text{eff} =
\tau_\text{bg} e^{\tau_\text{hold}/\tau_\text{bg}} (
e^{\tau_\text{PA}/\tau_\text{bg}} - 1 )$ accounts for the fact that
atoms are lost to one-body decay during the PA pulse with lifetime
$\tau_\text{bg}$ and that we hold the atoms for
$\tau_\text{hold}=\unit{100}{ms}$ between the end of the PA pulse and
the imaging. The number $\rho_{0}$ is the peak density after the PA
pulse and at large detuning. The definition of all other symbols
follows the main text. To make the fit numerically tractable, the
integral was approximated as a ten-term sum using Gauss-Laguerre
quadrature.

The thermally-averaged inelastic collision rate per particle
$\bar{K}_\text{in}$ was derived from the $s$-wave scattering matrix of
Bohn and Julienne, using the definition of the optical length
$\ell_\text{opt} = \Gamma_s / 2 k \gamma_m$ in terms of the stimulated
width $\Gamma_s$~\cite{om:bohn99,om:ciurylo05}. Here, we use a
thermally-averaged inelastic rate coefficient $\bar{K}_\text{in}
\equiv \langle \hbar k \sigma_\text{in} / \mu \rangle$, where
$\sigma_\text{in} = 2 \frac{\pi}{k^2} ( 1 - |S|^{2} )$~\cite{om:chin10}.
The angular brackets denote a thermal average over the initial
collision momenta. This average was performed on the inelastic rate
coefficient (rather than the density after the PA pulse) under the
assumption that prior to the PA pulse, many single-particle velocities
exist in every differential volume in the trap. Since single
collisions take place in a small volume, the spatial dependence of
$\bar{K}_\text{in}$ was carried through into the differential equation
for the density rather than averaged out.

To fit the inelastic loss spectra, $\bar{K}_\text{in}$ was expressed as
\begin{equation}
\label{om:eq:4}
\begin{split}
\bar{K}_\text{in}  (\Delta,u,\ell_\text{opt}\gamma_m,T) &=
\frac{8\sqrt{\pi} \hbar \ell_\text{opt}\gamma_m}{ \mu }
\int_{0}^{\infty} \!\!\! d\eta~\frac{\gamma \sqrt{\eta} \, \, e^{-\eta}}{
  D^2 + \Gamma^{2}/4}, \\
 D &\equiv \Delta + \frac{k_{B} T}{h} \eta - \nu_\text{rec} - \nu_{s} e^{-u/u_{0}},\\
 \Gamma &\equiv \gamma  + 2 k_\text{th} \ell_\text{opt} \gamma_m \sqrt{\eta},
\end{split}
\end{equation}
with dimensionless relative momentum magnitude $\eta\equiv
k^2/k_\text{th}^2$, thermal momentum $\hbar k_\text{th} = \sqrt{2\mu
  k_B T}$, center-of-mass PA photon recoil energy $h \nu_\text{rec}$,
trapping laser ac Stark shift at the center of the trap $\nu_s$, and
Planck's constant $h=2\pi\hbar$. The PA
laser with optical frequency $\nu_l$ is detuned from the PA resonance
by
\begin{equation}
  \label{om:eq:2}
  \Delta \equiv \nu_l - [\nu(\SSZ{}-\TPO{}) + \nu_n + \Delta\nu(I)],
\end{equation}
where $\nu(\SSZ{}-\TPO{})$ is the atomic transition frequency, and
$h\nu_n$ is the energy of state $n$ with respect to the free
threshold. The detuning term $\Delta\nu(I)$ accounts for the ac Stark
shift of the molecular resonance with respect to $I$.

The integral in Eq.~\ref{om:eq:4} was approximated as a 53-term sum using
Gauss-Laguerre quadrature. The quantities $(\ell_\text{opt}\gamma_m)$,
$T$, $u_{0}$, and a term added to the detuning to represent the line
center were allowed to vary. The parameter $T$ is used as a check
against the experimentally measured temperatures and agrees well with
the experiment. The Stark shift term $\nu_{s} e^{-u/u_{0}}$ was
included to account for the broadening of the atomic loss profile to
the blue side of a molecular resonance due to the ac Stark shift
induced by the trap. The molecular line width $\gamma$ was extracted
from the Monte-Carlo simulation in the next Section.

\section{Monte-Carlo Simulation ($2 k \ell_\text{opt}\gamma_m/\gamma \sim 1$)}
\label{sec:monte-carlo-simul}

To model the thermodynamic effects caused by the interplay of elastic
and inelastic collisions in an anharmonic trap with evaporation, we
use a Monte-Carlo simulation since analytic expressions such as those
presented in the previous Section are not available. The simulation is
based on classical particles moving in a conservative model potential
that includes the Gaussian beam shapes as well as gravity. A commonly
used method~\cite{om:wu96,om:wu97,om:goldwin05b} to include collisions in such
simulations is due to Bird~\cite{om:bird94} and detailed discussions of
the method in the context of ultracold atoms in optical traps can be
found in Refs.~\cite{om:gehm03,om:goldwin05}. The simulation procedure here
is described in more detail in Ref.~\cite{om:blatt11}.

The simulation uses the elastic and inelastic cross section formulas
derived in Section ``Scattering Matrix, Cross Sections, and Collision
Rates'':
\begin{equation}
  \label{om:eq:1}
  \begin{split}
    \sigma_\text{in}(k) & = \frac{4\pi}{k}
    \frac{\frac{\ell_\text{opt}\gamma_m}{\gamma}}
    {(\Delta+E/\hbar)^2/\gamma^2 + (1 + 2 k \frac{\ell_\text{opt}\gamma_m}{\gamma})^2/4}, \\
    \sigma_\text{el}(k) & \simeq 2k\frac{\ell_\text{opt}\gamma_m}{\gamma} \sigma_\text{in}(k), \\
  \end{split}
\end{equation}
where the second relation is an approximation valid in an ideal
ultracold gas like $^{88}$Sr where the background scattering length
$a_\text{bg}$ can be neglected. For a gas at non-zero temperature, the
detuning $\Delta$ includes the atomic motion and the trap ac Stark
shift as in the previous section.

The initial particle distribution is synthesized by dropping atoms
into the model potential and letting them evolve for several hundred
ms without collisions (using an embedded Runge-Kutta method). Each
particle is initially generated from independent Gaussian
distributions along the trap eigenaxes, both in position and velocity.
The initial widths of these Gaussian distributions are adjusted until
the particle distribution after having settled in the model potential
matches the experimental in-situ and TOF data when the PA laser is far
detuned from the OFR resonance.

To calibrate the Monte-Carlo simulation, we modelled a
three-dimensional isotropic harmonic trap of mK trap depth and checked
the thermodynamic effects of inelastic and elastic collisions
independently. Using a Gaussian initial phase-space density at several
$\mu$K and only allowing inelastic collisions at a
velocity-independent cross section, we recovered heating rates per
particle that match the corresponding analytic
expressions~\cite{om:blatt11}. Similarly, we check
cross-dimensional thermalization rates at a velocity-independent
elastic cross section. We found that the average number of elastic
collisions required for cross-dimensional thermalization in our
simulation agrees~\cite{om:blatt11} with the analytic
expressions from Ref.~\cite{om:goldwin05b}.

The Monte-Carlo simulation also includes the effect of scattering from
the atomic line. For the largest optical lengths in Fig.~4 of the main
paper, the atomic scattering rate is $\sim\unit{6}{s^{-1}}$
corresponding to 1.2 photons scattered per atom during the
\unit{200}{ms} exposure time. At these scattering rates, the photon
absorption along the PA laser direction and the random reemission
change the resulting cloud widths by less than 10\%. Due to the
spontaneous reemission as a spherical wave, the mean free path of a
scattered photon in a sample of OD~$\sim 2$ is still larger than the
typical cloud radii and we estimate that radiation trapping should
only introduce a small correction.

The final particle distribution is then imaged by randomly scattering
photons off of each particle and forming a histogram from the positions
of the scattering events over the image pulse time of \unit{50}{\mu
  s}, either in the trap or after TOF expansion for \unit{1.5}{ms}.
Random rescattering and corresponding velocity and position evolution
are included, and the position histogram is blurred to account for the
\unit{4}{\mu m} image resolution. The final image is then analyzed in
the same way as the experimental data to extract in-trap widths and
TOF temperatures. By adding the final measurement step instead of
calculating the covariance matrices of the position and velocity
distributions directly, the quantitative agreement with the experiment
was improved significantly in the regime of large inelastic losses.

\section{Maximal scattering length modification}

In the following, we build on the understanding from our current
experiment and try to estimate the maximum scattering length under the
most ideal conditions. We assume that atomic scattering is the
dominant loss mechanism, that $k\to 0$, and that there is no extra
molecular loss, such that $\gamma = 2 \gamma_a$. There is also no
spatial inhomogeneity, and thus it does not reflect the current
experimental conditions. Theoretical calculations of the molecular
line strength factors are used to make these estimates. Also note that
the line strength factors are only order-of-magnitude estimates beyond
$n<-5$.

\subsection{The Optically-Modified Scattering Length Constrained by Atomic Light Scatter}

In the $k \rightarrow 0$ limit, the scattering length change $\Delta
a$ due to an optical Feshbach resonance is given by
\begin{equation}
\label{eqn:delta_a}
  \Delta a = \ell_\text{opt} \frac{(\delta-\delta_{0}) \gamma_{m}}{(\delta-\delta_{0})^{2} + \gamma_{m}^{2}/4} =
  2 \ell_\text{opt} \frac{(\delta-\delta_{0}) \gamma_{a}}{(\delta-\delta_{0})^{2} + \gamma_{a}^{2}},
\end{equation}
where $\delta$ is $2 \pi$ times the detuning from atomic resonance,
$\delta_{0}$ is the difference between the molecular and atomic
resonance frequencies, $\gamma_{m}$ is the molecular linewidth,
$\gamma_{a}$ is the atomic linewidth, and $\gamma_{m} = 2 \gamma_{a}$.
The quantity $\ell_\text{opt} = \xi I$, where $I$ is the PA laser
intensity and $\xi \equiv \frac{\lambda_a^3}{16\pi c} \frac{|\langle n
  | E \rangle|^2}{k}$ is a resonance-specific constant that describes
the Sr+Sr molecular structure.

When using an OFR, $\delta$ must be chosen such that inelastic
collisional loss is small. In this regime, loss due to atomic light
scattering dominates. The atomic scattering rate $\Gamma_\text{sc}$ is
given by
\begin{equation}
\label{eqn:Gamma_sc}
  \Gamma_\text{sc} = \frac{\gamma_{a}}{2} \frac{s_{0}}{1 + s_{0} + 4 (\delta / \gamma_{a} )^{2}}.
\end{equation}
Here $s_{0} = I / I_\text{sat}$, where $I_\text{sat}$ is the saturation
intensity of the atomic transition. Typically the experiment
determines the maximum allowable value of $\Gamma_\text{sc}$, thereby
constraining $I$. Under this constraint, $I$ can be expressed as
\begin{equation}
\label{eqn:intensity}
  I = I_\text{sat} \frac{2 \Gamma_\text{sc} / \gamma_{a}}{1 - 2 \Gamma_\text{sc} /
    \gamma_{a}} \left( 1 + 4 \frac{\delta^{2}}{\gamma_{a}^{2}} \right)
  \approx 8 I_\text{sat} \left( \frac{\Gamma_\text{sc}}{\gamma_{a}} \right)
  \left( \frac{\delta}{\gamma_{a}} \right)^{2},
\end{equation}
where it has been assumed that $\gamma_{a} \gg \Gamma_\text{sc}$ and
$\delta \gg \gamma_{a}$, the latter of which is true due to
strontium's narrow intercombination line that is used in our work. The
maximum $\ell_\text{opt}$ for a given detuning is simply
\begin{equation}
\label{eqn:max_lopt}
  \ell_\text{opt} = \xi I = 8 \xi I_\text{sat} \left(
    \frac{\Gamma_\text{sc}}{\gamma_{a}} \right)
  \left( \frac{\delta}{\gamma_{a}} \right)^{2}.
\end{equation}

\begin{figure}[htb]
 \includegraphics[width=\columnwidth]{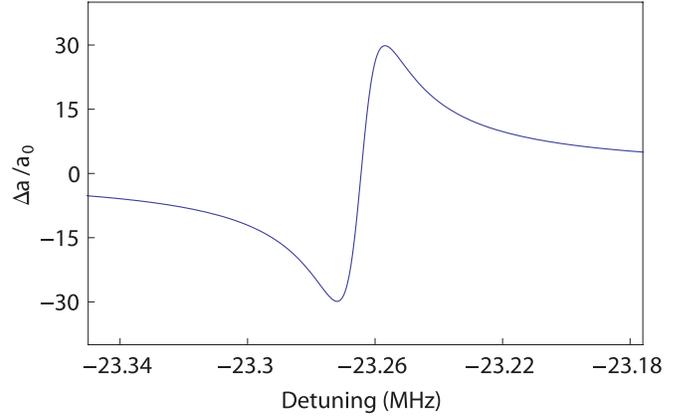}
 \caption{Equation~\ref{eqn:delta_a_constrained} plotted near
   molecular resonance. Here an $\xi$ corresponding to the
   \unit{-23.264}{MHz} line has been used, $\Gamma_\text{sc} =
   \unit{1}{s^{-1}}$, $\gamma_{a} = 2 \pi \times \unit{7.5}{kHz}$, and
   $I_\text{sat} =\unit{3}{\mu W/cm^2}$. This behavior is well
   approximated by $\Delta a_\text{dis}$ of
   Eq.~\ref{eqn:Delta_a_dis}.}
  \label{om:fig:dispersion}
\end{figure}

\begin{figure}[htb]
  \centering
  \includegraphics[width=\columnwidth]{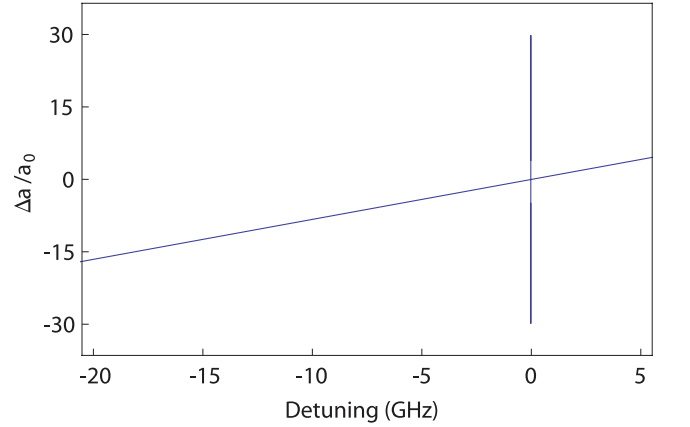}
  \caption{Equation~\ref{eqn:delta_a_constrained} plotted for large
    detunings. This plot was generated with the same values for $\xi$,
    $\delta_{0}$, $\Gamma_\text{sc}$, $\gamma_{a}$, and $I_\text{sat}$
    as those used in Fig.~\ref{om:fig:dispersion}. This behavior of
    $\Delta a$ for large detunings is well approximated by $\Delta
    a_\text{lin}$ of Eq.~\ref{eqn:Delta_a_lin}. As will be discussed
    in Section ``The Large Detuning Case,'' these detunings are
    unphysical given the molecular structure and the coupled-channnels
    theory discussed in the main text.}
  \label{om:fig:linear}
\end{figure}

Thus $\Delta a$, constrained to a certain $\Gamma_\text{sc}$, is
\begin{equation}
\label{eqn:delta_a_constrained}
  \Delta a = 16 \xi I_\text{sat} \left( \frac{\Gamma_\text{sc}}{\gamma_{a}}
  \right)
  \frac{\delta^{2}}{\gamma_{a}^{2}} \frac{(\delta - \delta_{0})\gamma_{a}}
  {(\delta - \delta_{0})^{2} + \gamma_{a}^{2}}.
\end{equation}
This expression is plotted in Figs.~\ref{om:fig:dispersion} and
\ref{om:fig:linear}. Within tens of linewidths $\gamma_{m}$ of $\delta =
\delta_{0}$, the factor $\delta^{2}$ in the numerator of
Eq.~\ref{eqn:delta_a_constrained} is slowly varying and can be
approximated as $\delta_{0}^{2}$. Let $\Delta a_\text{dis}$ refer to
$\Delta a$ in this regime. Therefore,
\begin{equation}
\label{eqn:Delta_a_dis}
  \Delta a_\text{dis} = 16 \xi I_\text{sat} \left(
    \frac{\Gamma_\text{sc}}{\gamma_{a}} \right)
  \frac{\delta_{0}^{2}}{\gamma_{a}^{2}} \frac{(\delta -
    \delta_{0})\gamma_{a}}
  {(\delta - \delta_{0})^{2} + \gamma_{a}^{2}} ,
\end{equation}
which has the dispersion shape of Fig.~\ref{om:fig:dispersion}. When
$\delta \gg \delta_{0}$, $\left| \Delta a \right|$ varies linearly
with $\delta$. Let $\Delta a_\text{lin}$ be the expression for $\Delta a$
when $\delta \gg \delta_{0}$.
\begin{equation}
\label{eqn:Delta_a_lin}
  \Delta a_\text{lin} = 16 \xi I_\text{sat}
  \left(\frac{\Gamma_\text{sc}}{\gamma_{a}} \right)
  \left( \frac{\delta}{\gamma_{a}} \right),
\end{equation}
which describes the linear behavior evident in Fig.~\ref{om:fig:linear}.

\subsection{The Maximum Useful Scattering Length Near a Molecular Resonance}
\label{sec:max_a}

\begin{figure}[htb]
  \centering
  \includegraphics[width=\columnwidth]{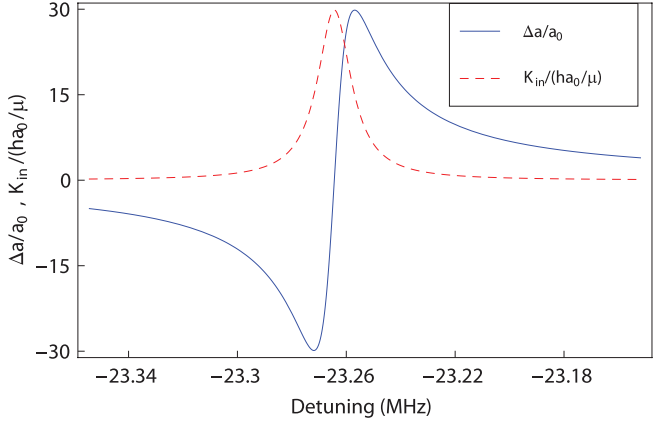}
  \caption{Equation~\ref{eqn:delta_a_constrained} plotted in blue and
    Eq.~\ref{eqn:Kin} plotted in red. This plot was generated
    with the same values for $\xi$, $\delta_{0}$, $\Gamma_\text{sc}$,
    $\gamma_{a}$, and $I_\text{sat}$ as those used in
    Fig.~\ref{om:fig:dispersion}. Note that $K_\text{in}$ is significant near a
    molecular resonance but that this quantity drops off faster than
    $\Delta a$.}
  \label{om:fig:comparison}
\end{figure}

Near a molecular resonance where $|\Delta a|$ has dispersive behavior,
it follows from Eq.~\ref{eqn:Delta_a_dis} that $|\Delta a|$ has a
local maximum at $|\delta - \delta_{0}| \approx \pm \gamma_{a}$.
However, at these small detunings inelastic collisional loss must be
considered. For $k \rightarrow 0$, the inelastic rate coefficient
$K_\text{in}$ is given by
\begin{equation}
\label{eqn:Kin}
  \begin{split}
  K_\text{in} &= \frac{4\pi \hbar \ell_\text{opt}}{\mu}
  \frac{\gamma_{m}^{2}}{(\delta-\delta_{0})^{2} + \gamma_{m}^{2}/4}\\
  &= \frac{16\pi \hbar \ell_\text{opt}}{\mu}
  \frac{\gamma_{a}^{2}}{(\delta-\delta_{0})^{2} + \gamma_{a}^{2}}.\\
  \end{split}
\end{equation}
Putting Eq.~\ref{eqn:max_lopt} into Eq.~\ref{eqn:Kin}, it
can easily be shown that the maximum value of $K_\text{in}$, denoted by
$K^\text{max}_\text{in}$, occurs at $\delta \approx \delta_{0}$, consistent with the
case when $\ell_\text{opt}$ in Eq.~\ref{eqn:delta_a} is not
constrained by atomic light scatter.

As Fig.~\ref{om:fig:comparison} shows, inelastic loss (described by
$K_\text{in}$) is significant near $\delta_{0}$. Nevertheless, this
figure makes it apparent that one can choose a detuning that is just
outside of the influence of $K_\text{in}$ but that still yields a
significant $\Delta a$. This detuning is implied upon setting
$K_\text{in} = \kappa K^\text{max}_\text{in}$, where $\kappa$
describes the fraction of molecular loss deemed acceptable by the
experiment. Solving this equation for detuning yields that $\delta -
\delta_{0} \approx \pm \gamma_{a}/\sqrt{\kappa}$. Evaluating
Eq.~\ref{eqn:delta_a_constrained} at this detuning produces $\Delta
a_\text{max}$,
\begin{equation}
\label{eqn:Delta_a_max}
  \Delta a _\text{max} \approx \pm 16 \xi I_\text{sat} \sqrt{\kappa}
  \left( \frac{\Gamma_\text{sc}}{\gamma_{a}} \right)
  \left( \frac{\delta_{0}}{\gamma_{a}} \right)^{2},
\end{equation}
where $\Delta a_\text{max}$ is the maximum scattering length change when
$\Delta a$ is dispersive and inelastic loss and atomic light scatter
are negligible. The intensity corresponding to $\Delta a_\text{max}$, given
by Eq.~\ref{eqn:intensity}, is
\begin{equation}
\label{eqn:intensity_at_max}
  I_{\Delta a_\text{max}} \approx 8 I_\text{sat} \left(
    \frac{\Gamma_\text{sc}}{\gamma_{a}} \right)
  \left( \frac{\delta_{0}}{\gamma_{a}} \right)^{2}.
\end{equation}

\subsection{The Large-Detuning Case}
\label{sec:large_delta_case}

Although $|\Delta a|$ in Eq.~\ref{eqn:delta_a_constrained} seems to
increase without bound when $\delta \gg \delta_{0}$, coupled-channels
OFR theory (which so far has not been considered in this discussion)
dictates that if $|\Delta a|$ is based on the $n$th resonance, the
modification to the scattering length will vanish at some point before
$\delta$ is equal to the detuning of the $(n-1)$th resonance. See the
main text for a further discussion of this coupled-channels effect.
One must consider whether the molecular detuning $\delta-\delta_0$ can
be sufficiently large to take advantage of the linear increase with
$\delta$ but not so far that coupled-channels effects become a
concern.

Let $\delta_\text{lin}$ be the detuning at which the magnitude of
$\Delta a_\text{lin}$ is equal to $|\Delta a_\text{max}|$. It follows
that $|\delta_\text{lin}| = \sqrt{\kappa}
\delta_{0}^{2}/\gamma_a$. If $\delta_\text{lin}$ is comfortably
outside the regime where coupled-channels effects are a concern, then
the maximum useful scattering length occurs when one is many
linewidths $\gamma_{m}$ detuned from $\delta_{0}$ but not far enough
detuned to be close to other resonances. However, theoretical values
for $\delta_{0}$ make it clear that detuning to $\delta_\text{lin}$
would always require crossing another molecular resonance, so
coupled-channels effects dictate that the maximum scattering length
one can achieve is given by $\Delta a_\text{max}$
(Eq.~\ref{eqn:Delta_a_max}), occurring when $\delta$ is as close as
possible to a molecular line, while molecular losses are constrained
to a given level. Table~\ref{tab:1} considers 6 different molecular
resonances and their associated values of $\delta_\text{lin}$, which
are many orders of magnitude larger than $\delta_{0}$ for subsequent
resonances.

\begin{table}
\begin{center}
\begin{tabular}{r|cc}
  $n$~ & $\delta_{0}$ (GHz) & $\delta_\text{lin}$ (GHz)\\
  \hline
  -2 & -0.023 & -7.22 \\
  -3 & -0.221 & -655 \\
  -4 & -1.084 & -1.57 $\times 10^{4}$ \\
  -5 & -3.460 & -1.60 $\times 10^{5}$ \\
  -6 & -8.400 & -9.40 $\times 10^{5}$ \\
  -7 & -17.0 & -3.84 $\times 10^{6}$ \\
  % -8 & -30.4 & -1.23 $\times 10^{7}$ \\
  % -9 & -49.7 & -3.30 $\times 10^{7}$ \\
  % -10 & -76.1 & -7.73 $\times 10^{7}$ \\
  % -11 & -110 & -1.64 $\times 10^{8}$ \\
  % -12 & -154 & -3.20 $\times 10^{8}$ \\
  % -13 & -209 & -5.85 $\times 10^{8}$ \\
  % -14 & -275 & -1.01 $\times 10^{9}$ \\
  % -15 & -355 & -1.68 $\times 10^{9}$ \\
  % -16 & -448 & -2.68 $\times 10^{9}$ \\
\end{tabular}
\end{center}
\caption{Theoretical values for $\delta_{0}$ for the $n=-2$ through -7
  $0_u$ resonances and the associated values for $\delta_\text{lin}$. Here
  $\kappa = 0.01$.}
\label{tab:1}
\end{table}

\subsection{Conclusion}

\begin{table}
\begin{center}
\begin{tabular}{r|cccc}
  $n$~ & $\delta_{0}$ (GHz) & $\xi$ ($\mathrm{a_{0}/(W/cm^{2}})$) & $| \Delta a_\text{max} |$ ($a_{0}$) & $I_{\Delta a_\text{max}}$ ($\mathrm{W/cm^{2}}$)\\
  \hline
  -2 & -0.023 & 6110 & 5.97 & 4.89 $\times 10^{-3}$ \\
  -3 & -0.221 & 32.8 & 2.92 & 0.443 \\
  -4 & -1.084 & 27.9 & 59.3 & 10.6 \\
  -5 & -3.460 & 3.30 & 71.5 & 108 \\
  -6 & -8.400 & 0.272 & 34.7 & 637 \\
  -7 & -17.0 & 0.012 & 6.21 & 2.61 $\times 10^{3}$ \\
  % -8 & -30.4 & 2.71 $\times 10^{-5}$ & 0.045 & 8.34 $\times 10^{3}$ \\
  % -9 & -49.7 & 6.72 $\times 10^{-4}$ & 3.00 & 2.23 $\times 10^{4}$ \\
  % -10 & -76.1 & 4.87 $\times 10^{-4}$ & 5.10 & 5.24 $\times 10^{4}$ \\
  % -11 & -110 & 2.24 $\times 10^{-4}$ & 4.97 & 1.11 $\times 10^{5}$ \\
  % -12 & -154 & 9.69 $\times 10^{-5}$ & 4.20 & 2.17 $\times 10^{5}$ \\
  % -13 & -209 & 4.91 $\times 10^{-5}$ & 3.90 & 3.97 $\times 10^{5}$ \\
  % -14 & -275 & 3.29 $\times 10^{-5}$ & 4.52 & 6.88 $\times 10^{5}$ \\
  % -15 & -355 & 2.85 $\times 10^{-5}$ & 6.50 & 1.13 $\times 10^{6}$ \\
  % -16 & -448 & 2.90 $\times 10^{-5}$ & 10.5 & 1.81 $\times 10^{6}$ \\
\end{tabular}
\end{center}
\caption{Theoretical values for $\delta_{0}$ and $\xi$ for the $n=-2$
  through -7 $0_u$ resonances are used to calculate the maximum useful
  scattering length. The intensities required for these scattering
  lengths are also included. The optimal detuning for each resonance
  is given by  $|\delta - \delta_{0}| = \gamma_{a}/\sqrt{\kappa} = 10
  \gamma$
  for the value $\kappa = 0.01$ that was used in the table.}
\label{tab:2}
\end{table}

In light of Section ``The Large-Detuning Case'', we conclude that for
a given molecular line, the best optically-modified scattering length
occurs for a laser detuned just outside the influence of inelastic
loss and with an intensity just low enough to prevent heating due to
atomic light scatter. This optimal detuning is given by $\delta -
\delta_{0} = \gamma_{a} / \sqrt{\kappa}$. Table~\ref{tab:2} lists
theoretical $0_u$ values for $\xi$ and $\delta_{0}$ as well the
associated values for $\Delta a_\text{max}$ and $I_{\Delta
  a_\text{max}}$. Values of $\Gamma_\text{sc} = \unit{1}{s^{-1}}$ and
$\kappa = 0.01$ have been used. For the strontium \SSZ{}-\TPO{} line,
$\gamma_{a} = 2 \pi \times \unit{7.5}{kHz}$ and $I_\text{sat} =
\unit{3}{\mu W/cm^{2}}$. These numbers yield an optimal detuning of
$\delta - \delta_{0} = 2 \pi \times \unit{75}{kHz}$. The intensity
column was included to point out that many resonances cannot be used
on technical grounds (for instance intensities greater than a few
kW/cm$^2$ are impractical for external cavity diode lasers).

Under the ideal conditions of zero temperature, no spatial
inhomogeneity, no extra molecular loss and the assumption that atomic
light scattering is the dominant loss mechanism, Table~\ref{tab:2}
shows that the resonance location of the $n = -5$ line is the best
compromise between a large detuning from atomic resonance (which
allows for more PA laser intensity without atomic light scatter) and
the fact that $\xi$ rapidly decreases with increasing $|\delta|$.
According to Eq.~\ref{eqn:Delta_a_max}, only moderate gains in $\Delta
a_\text{max}$ can be achieved by allowing for more inelastic loss
(when detuning closer to a molecular line) since any increase in
$K_\text{in}$ will only be accompanied by a $\sqrt{K_\text{in}}$
increase in $|\Delta a_\text{max}|$. More significant gains can be
achieved by shortening experimental time scales to allow for larger
values of $\Gamma_\text{sc}$ since $|\Delta a_\text{max}|$ is linear
in $\Gamma_\text{sc}$.

\end{document}